\documentclass[twocolumn,english,pra,aps,superscriptaddress]{revtex4-1}
\usepackage[T1]{fontenc}
\usepackage{color}
\usepackage{babel}
\usepackage{mathtools}
\usepackage{bm}
\usepackage{graphicx}
\usepackage{esint}
\usepackage{amsfonts}
\usepackage{graphicx}
\usepackage{float}
\usepackage{subfigure}
\usepackage{amsthm}

\usepackage[usenames,dvipsnames,svgnames,table]{xcolor}
\usepackage[unicode=true,pdfusetitle,
bookmarks=true,bookmarksnumbered=false,bookmarksopen=false,
breaklinks=true,pdfborder={0 0 0},backref=false,colorlinks=true]{hyperref}
\hypersetup{linkcolor=NavyBlue,urlcolor=NavyBlue,citecolor=NavyBlue}
\usepackage{breakurl}

\begin{document}
\title {Quantum metrology with precision reaching beyond-$1/N$ scaling through
$N$-probe entanglement generating interactions}

\author{Xing Deng}
\affiliation{MOE Key Laboratory of Fundamental Physical Quantities Measurement,
Hubei Key Laboratory of Gravitation and Quantum Physics,
PGMF and School of Physics, Huazhong University of Science and Technology,
Wuhan 430074, People's Republic of China}

\author{Shou-Long Chen}
\affiliation{MOE Key Laboratory of Fundamental Physical Quantities Measurement,
Hubei Key Laboratory of Gravitation and Quantum Physics,
PGMF and School of Physics, Huazhong University of Science and Technology,
Wuhan 430074, People's Republic of China}

\author{Mao Zhang}
\affiliation{MOE Key Laboratory of Fundamental Physical Quantities Measurement,
Hubei Key Laboratory of Gravitation and Quantum Physics,
PGMF and School of Physics, Huazhong University of Science and Technology,
Wuhan 430074, People's Republic of China}

\author{Xiao-Fan Xu}
\email{xiaofanxu@live.com}
\affiliation{China Satellite Network Innovation Co., Ltd., Beijing 100029,
People's Republic of China}

\author{Jing Liu}
\email{liujingphys@hust.edu.cn}
\affiliation{MOE Key Laboratory of Fundamental Physical Quantities Measurement,
 Hubei Key Laboratory of Gravitation and Quantum Physics,
PGMF and School of Physics, Huazhong University of Science and Technology,
Wuhan 430074, People's Republic of China}

\author{Zhi Gao}
\affiliation{MOE Key Laboratory of Fundamental Physical Quantities Measurement,
Hubei Key Laboratory of Gravitation and Quantum Physics,
PGMF and School of Physics, Huazhong University of Science and Technology,
Wuhan 430074, People's Republic of China}

\author{Xiao-Chun Duan}
\affiliation{MOE Key Laboratory of Fundamental Physical Quantities Measurement,
Hubei Key Laboratory of Gravitation and Quantum Physics,
PGMF and School of Physics, Huazhong University of Science and Technology,
Wuhan 430074, People's Republic of China}

\author{Min-Kang Zhou}
\affiliation{MOE Key Laboratory of Fundamental Physical Quantities Measurement,
Hubei Key Laboratory of Gravitation and Quantum Physics,
PGMF and School of Physics, Huazhong University of Science and Technology,
Wuhan 430074, People's Republic of China}

\author{Lushuai Cao}
\email{lushuai\_cao@hust.edu.cn}
\affiliation{MOE Key Laboratory of Fundamental Physical Quantities Measurement,
 Hubei Key Laboratory of Gravitation and Quantum Physics,
PGMF and School of Physics, Huazhong University of Science and Technology,
Wuhan 430074, People's Republic of China}

\author{Zhong-Kun Hu}
\email{zkhu@hust.edu.cn}
\affiliation{MOE Key Laboratory of Fundamental Physical Quantities Measurement,
 Hubei Key Laboratory of Gravitation and Quantum Physics,
PGMF and School of Physics, Huazhong University of Science and Technology,
Wuhan 430074, People's Republic of China}

\begin{abstract}
Nonlinear interactions are recognized as potential resources for quantum metrology,
facilitating parameter estimation precisions that scale as the exponential Heisenberg
limit of $2^{-N}$. We explore such nonlinearity and propose an associated quantum
measurement scenario based on the nonlinear interaction of $N$-probe entanglement
generating form. This scenario provides an enhanced precision scaling of $D^{-N}/(N-1)!$
with $D > 2$ a tunable parameter. In addition, it can be readily implemented in
a variety of experimental platforms and applied to measurements of a wide range
of quantities, including local gravitational acceleration $g$, magnetic field,
and its higher-order gradients.
\end{abstract}

\maketitle

\section{Introduction}
Quantum metrology aims at improving the practical measurement precisions via
quantum advantages. Various measurement scenarios have been
proposed~\cite{Giovannetti2006,Braun2018,Giovannetti2011},
capable of achieving precisions beyond the standard quantum limit (SQL) of $1/\sqrt{N}$,
or even approach the Heisenberg limit $1/N$ for an ensemble of N particles, with
nonclassical probe states, such as the entangled~\cite{Giovannetti2004,Mitchell2004,zou2018,
luo2017deterministic} and squeezed states~\cite{wineland1994squeezed,Gross2012,Bao2020},
as probe states to interferometers. Besides the nonclassical probe states, quantum
parameterization processes based on nonlinear interactions between the probes and
the to-be-measured (TBM) system, can also improve the precision by amplifying the
phase imprinted by the TBM quantity to the probes. For instance, the nonlinear
$k$-body interaction can reach a precision scaling of $N^{-k}$ ($N^{-k+\frac{1}{2}}$),
with (without) entanglement of the probe~\cite{PhysRevA.72.045801,PhysRevLett.98.090401,
PhysRevA.77.053613,Napolitano2011interaction-based,NIE2018469}, and the nonlinear
$N$-body entanglement generating interaction can even achieve an exponential
Heisenberg limit of $2^{-N}$ for measuring its interaction strength~\cite{roy2008exponentially}.
These studies support higher than SQL scaling with $N$ for the precision and are
broadly applicable.

Despite that the quantum measurement schemes based on the nonlinear interactions
can provide the precision with the beyond-$1/N$ scaling, they suffer from a common
constraint of weak applicability, arising from the facts that, the nonlinear
interaction coupling the TBM system with the probes is hard to engineer, and more
severely, the enhanced precision scaling only works for the measurement of the
nonlinear interaction strength, which cannot be assigned to an arbitrary TBM quantity.
In order to circumvent the constraint, we propose a new nonlinear measurement scheme,
which can generalize the beyond-$1/N$ precision scaling to measurements of a wider range
of quantities. Instead of introducing a nonlinear interaction between the probes and
the TBM system, this measurement scheme deploys the $N$-body entanglement generating
interaction to locally couple $N$ probes, and we will simply call it $N$-probe
entanglement generation (NPEG) interaction scheme. It will be demonstrated that,
for one thing, this NPEG-based scheme maintains the beyond-$1/N$ precision scaling,
and more importantly, once the NPEG interaction is engineered, the nonlinearly coupled
probes can be associated to various TBM systems and realize the measurements
of different TBM quantities with the improved precision.

The engineering of the NPEG interaction can benefit from the highly developed
investigations on effective Hamiltonian engineering. Earlier studies on
effective Hamiltonian engineering have been carried out on various
platforms, especially in ultracold atomic ensembles for generating effective
coupling channels between intrinsically isolated states and tuning the coupling
strength. The effective coupling channels are usually engineered via
either applying external couplings through lasers~\cite{PhysRev.175.453,
RevModPhys.76.1037,PhysRevX.4.031027,RevModPhys.89.011004}, or by
tailoring higher order processes intermediated by energetically detuned
states~\cite{PhysRevLett.85.3987,PhysRevLett.89.090401,dai2017four-body,Brown540}.
Particularly, nonlinear interactions have been proposed and even experimentally
realized~\cite{PhysRevLett.89.090401,paredes2008minimum,nascimbene2012experimental,
Brown540,dai2017four-body} for the investigations of e.g. quantum simulations,
which could be directly applied for the NPEG-based measurements. The studies on
effective Hamiltonian engineering help to lay solid foundation to the NPEG-based
measurements, which in turn provide new applications to the engineering studies.

This paper is organized as follows: in section II the NPEG-based measurement scheme
is introduced, with an intuitive illustration on the precision scaling; in section
III the performance of the NPEG-based measurement scheme, in terms of the precision
scaling, the time resource, the robustness, is analyzed with the quantum and
classical Fisher informations; in section IV, the applicability of the scheme is
investigated with a variety of experimentally realizable setups of the measurement
scheme; a summary is present in section V.

\section{NPEG~measurement~scheme}
The key element of the proposed measurement scheme is the NPEG nonlinear interaction,
which can be engineered through higher-order processes on various setups~\cite{PhysRev.175.453,
RevModPhys.76.1037,PhysRevX.4.031027,RevModPhys.89.011004,PhysRevLett.85.3987,
PhysRevLett.89.090401,PhysRevLett.95.040402,paredes2008minimum,nascimbene2012experimental,
Brown540,dai2017four-body}. We briefly illustrate the engineering procedure with
the example of a bosonic system composed of $N$ identical bosons, where each
boson can only occupy two states, denoted by $|a\rangle$ and $|b\rangle$. The
Hamiltonian of the original system is
\begin{equation} \label{ham0}
H_\mathrm{org}=
\frac{U}{2}(\hat{a}^\dag\hat{a}^\dag\hat{a}\hat{a}+\hat{b}^\dag\hat{b}^\dag\hat{b}\hat{b})
+j(\hat{a}^\dag \hat{b}+\hat{a} \hat{b}^{\dagger})-\frac{\delta_\mathrm{0}}{2}(\hat{n}_a-\hat{n}_b),
\end{equation}
in which $\hat{a}/\hat{b}$ ($\hat{a}^{\dag}/\hat{b}^{\dag}$) annihilates (creates)
a boson in the state $|a\rangle/|b\rangle$, and $\hat{n}_{a}=\hat{a}^{\dag}\hat{a}$
($\hat{n}_{b}=\hat{b}^{\dag}\hat{b}$). In the strong interaction regime with
$j\ll U$, the interaction of bosons divides the Hilbert space into a set of
subspaces, each of which is composed of energetically resonant basis states of
fixed integer number of atoms $N=n_\mathrm{a}+n_\mathrm{b}$. The states
$\{|N, 0\rangle\!:=|\!\!\Uparrow\rangle, \,|0, N\rangle\!:=|\!\!\Downarrow\rangle\}$
span such a subspace $\mathbb{S}$,
which denote states of all bosons confined in the mode $|a\rangle$ and
$|b\rangle$, respectively. Within $\mathbb{S}$, $|\!\!\Uparrow\rangle$ and
$|\!\!\Downarrow\rangle$ are not directly coupled by $H_{\mathrm{org}}$,
but an effective coupling between the two states can be engineered
through the higher-order process induced by the single-particle hopping $j$.
The resulting effective Hamiltonian (see appendix~\ref{Heff} for details)
within $\mathbb{S}$ reads
\begin{equation} \label{eq:ham_eff}
H_\mathrm{eff}\!=\!J_\mathrm{eff}\left(|\!\!\Uparrow\rangle\langle\Downarrow\!\!|
+|\!\!\Downarrow\rangle\langle\Uparrow\!\!|\right)
-\frac{1}{2}\Delta_\mathrm{0}\left(|\!\!\Uparrow\rangle\langle\Uparrow\!\!|
-|\!\!\Downarrow\rangle\langle\Downarrow\!\!|\right),
\end{equation}
in which the first term is the effective coupling of the nonlinear NPEG form,
with $J_{\mathrm{eff}}=\frac{NU}{D^N(N-1)!}$ $(D=U/j)$, and the second term is
induced by the single-particle detuning with $\Delta_\mathrm{0}=N \delta_\mathrm{0}$.
While such NPEG interactions have been widely investigated in various
fields~\cite{S.F2007,Dai2016,PhysRevLett.85.3987,dai2017four-body,PhysRevLett.89.090401},
for the purpose of e.g. quantum simulations, here we propose that they are also
helpful to quantum measurements with improved precision.

\begin{figure}[tp]
\includegraphics[width=8.5cm]{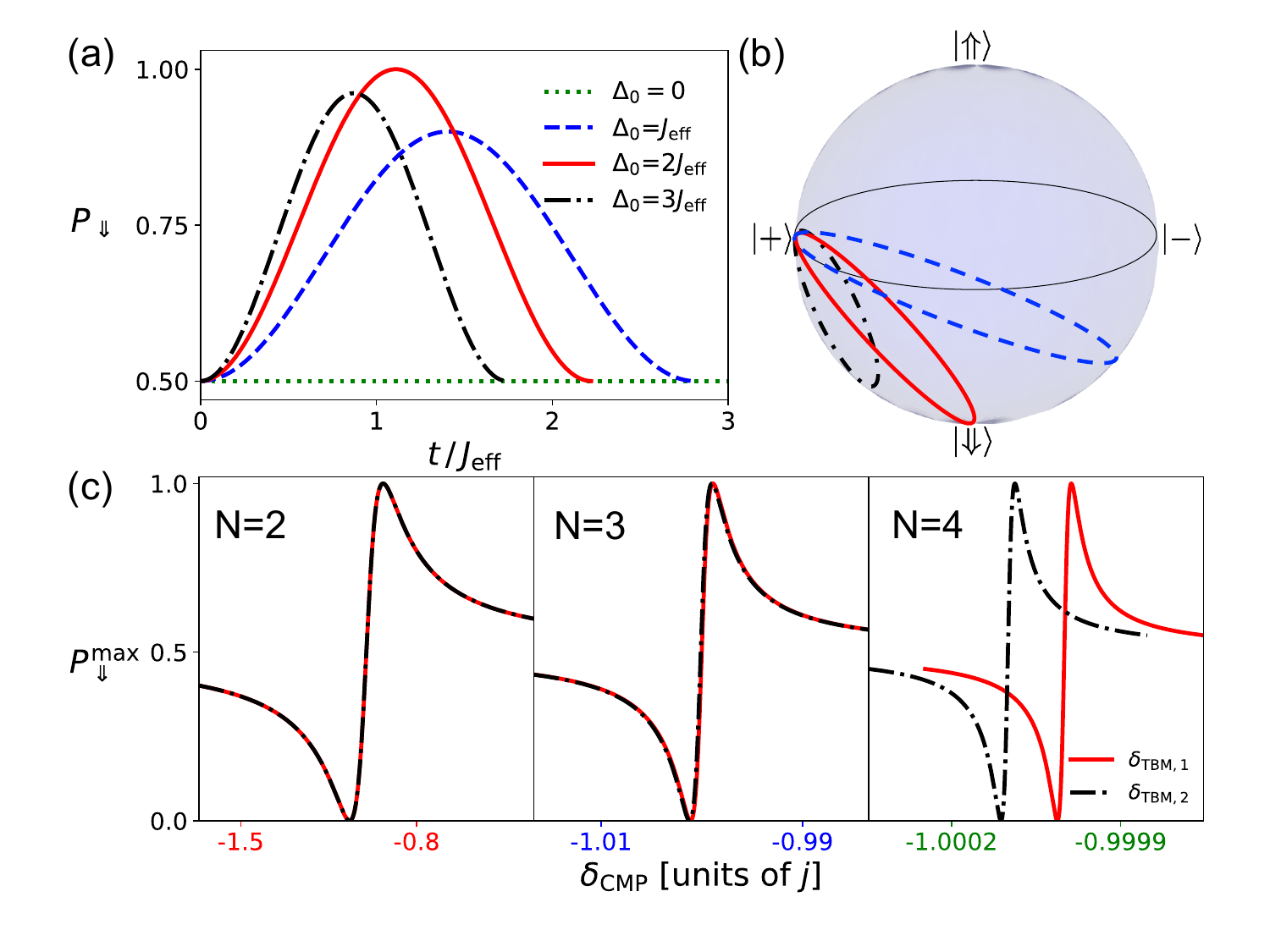}
\caption{(a) Time evolution of the probability $P_{\Downarrow}$ with the
initial state $|+\rangle$ for the total detuning $\Delta_\mathrm{0}=0$
(dotted green line), $\Delta_\mathrm{0}=J_{\mathrm{eff}}$ (dashed blue line),
$\Delta_\mathrm{0}=2J_{\mathrm{eff}}$ (solid red line) and
$\Delta_\mathrm{0}=3J_{\mathrm{eff}}$ (dash-dotted black line).
(b) Schematic of evolution trajectories for the initial state $|+\rangle$ with
the values of $\Delta_\mathrm{0}$ discussed in (a) (lines of the same color
correspond to the same value of $\Delta_\mathrm{0}$). (c) The maximum tunneling
amplitude $P^{\max}_\mathrm{\Downarrow}$ as a function of $\delta_{\mathrm{CMP}}$
under slightly different values of $\delta_\mathrm{TBM,1}$ and
$(1+10^{-4})\delta_\mathrm{TBM,1}$ for $N=2$ (left panel), $N=3$ (middle panel)
and $N=4$ (right panel). Here $\delta_{\mathrm{TBM,1}}$ is assumed to take the
same value of $j$ ($j$ is set to be 1).}
\label{fig:fig1}
\end{figure}

The NPEG-based measurement scheme is based on the dynamical process of the
many-body correlation induced tunneling (MBCIT), which is a special type of
correlated tunneling generalized from the single-particle correlation induced
tunneling ~\cite{refId0,CAO2017303}. During the dynamical process, the system
periodically oscillate between the initial and the corresponding target, for
instance between the state $|+\rangle:=(|\!\!\Uparrow\rangle+|\!\!\Downarrow\rangle)/\sqrt{2}$
and $|\!\!\Downarrow\rangle$. Defining $\Gamma:=\Delta_\mathrm{0}/J_\mathrm{eff}$,
the probability to be in the state $|\!\!\Downarrow\rangle$ during MBCIT from
the initial state of $|+\rangle$ is
\begin{equation} \label{Pmax}
P_\mathrm{\Downarrow}(t)=\frac{1}{2}+\frac{\Gamma-\Gamma\cos(\omega t)}{4+\Gamma^2},
\end{equation}
with $\omega=\sqrt{4J_\mathrm{eff}^2+\Delta_\mathrm{0}^2}$.
The maximum tunneling amplitude $P^{\max}_\mathrm{\Downarrow}
=\frac{(2+\Gamma)^2}{2(4+\Gamma^2)}$ is reached at the half period,
which builds up a dependence between $\Delta_\mathrm{0}$
and the experimentally accessible observable $P^{\max}_\mathrm{\Downarrow}$.
Such dependence permits the estimation of $\Delta_\mathrm{0}$, and
equivalently $\delta_\mathrm{0}$, through the measurement of $P^{\max}_\mathrm{\Downarrow}$.
The derivative $\partial_{\Delta_\mathrm{0}} P^{\max}_{\mathrm{\Downarrow}}$
determines the precision of the estimation, and shows a beyond-exponential dependence
on the number of bosons $N$, implying a beyond-$1/N$ scaling
of the estimation precision.
In Fig.~\ref{fig:fig1}(a), the temporal evolutions of $P_\mathrm{\Downarrow}(t)$
are depicted for different detunings, and the dependence of the tunneling amplitude on
the detuning is illustrated. The MBCIT can also be viewed as an effective precession
on the Bloch sphere spanned by $\{|\!\!\Downarrow\rangle,|\!\!\Uparrow\rangle\}$,
as shown in Fig.~\ref{fig:fig1}(b), in which
$H_{\mathrm{eff}}$ plays the role of a biased magnetic field.

The NPEG-based measurement scheme takes each boson in the two-mode system
as a probe for measuring the detuning (of some TBM system), specified as
$\delta_\mathrm{TBM}$ in the following. Benefiting  from the sharp particle-number
dependence of $\partial_{\Delta_\mathrm{0}}P^{\max}_{\mathrm{\Downarrow}}$
in the MBCIT process, a precision beyond-$1/N$ scaling is achieved.
Further optimization of the scheme with respect to the value of the detuning
itself can also affect the precision, and this optimization is carried out by
introducing a compensating of the single-particle detuning, with strength
$\delta_{\mathrm{CMP}}$, to maximize the measurement precision of $\delta_\mathrm{TBM}$.
The total single-particle and many-body detunings, namely $\delta_\mathrm{0}$ and
$\Delta_\mathrm{0}$, respectively, are then composed of the TBM and the
compensate ones, and the TBM detuning can be estimated by
$\delta_{\mathrm{TBM}}=\delta_\mathrm{0}-\delta_{\mathrm{CMP}}$,
where the total detuning is deduced from the measured $P^{\max}_\mathrm{\Downarrow}$.
Provided that the value of $\delta_{\mathrm{CMP}}$ can be obtained
with a high enough precision, the precision of $\delta_{\mathrm{TBM}}$ is
determined also to a high precision by that of $P^{\max}_\mathrm{\Downarrow}$,
which can be optimized by tuning $\delta_{\mathrm{CMP}}$.

For a fixed $\delta_{\mathrm{TBM}}$, the $\delta_{\mathrm{CMP}}$
is scanned and the corresponding $P^{\max}_\mathrm{\Downarrow}$ is measured, from which
the total detuning with the highest precision is deduced.
Figure.~\ref{fig:fig1}(c) plots the response of
$P^{\max}_\mathrm{\Downarrow}$ to $\delta_{\mathrm{CMP}}$ in the presence of two
slightly different TBM detunings $\delta_{\mathrm{TBM,1}}$ (red solid lines)
and $\delta_{\mathrm{TBM,2}}=(1+10^{-4})\delta_{\mathrm{TBM,1}}$
(black dashed lines) for different probe number of
$N=2$ (left panel), $N=3$ (middle panel), and $N=4$ (right panel).
As calibrated by the colored tick-marks on the $x$-axis, the response grows sharply
as $N$ increases, indicating that the NPEG measurement scheme indeed provides a
significant precision improvement with increasing probe number. Meanwhile, it
shows that only with $N=4$ probes can distinguish the two slightly different TBM
detunings via statistics from the measured state populations, indicating that the
NPEG-based scheme can also enhance measurement resolution. The precision and the
resolution with enhanced scaling over the number of probes will benefit realistic
metrological scenarios such as gravity and its gradient.

The NPEG-based measurement described above is designed based on a
different mechanism from the proposal in Ref.~\cite{roy2008exponentially}
and provides enhanced precision scaling as well as a broader applicability.
Since the uncertainty of parameter estimation can be determined by
$\Delta\delta=\Delta P/|\partial_{\delta} P|$, where $\Delta \delta$ and
$\Delta P$ refer to the uncertainties of the estimated parameter $\delta$ and
the experimentally measured probability $P$, respectively, the NPEG-based
scheme can simultaneously improve $\Delta P$ and $|\partial_{\delta} P|$,
and consequently provides a new precision scaling, as will be
demonstrated below.
Moreover, in the NPEG-based scheme, once the nonlinear interaction between the
probes is realized, the nonlinearly coupled probes can be subjected to different
TBM systems and measure different quantities,
enabling broad practical applicability of the nonlinear interactions as the
quantum resource.

\section{Optimization and characterization}
The precision of the NPEG-based measurement can be analyzed and
optimized according to the Quantum Cram\'{e}r-Rao bound~\cite{Helstrom1976,Holevo1982},
one of the most well-studied bounds in quantum metrology for both single-~\cite{Paris2009,Toth2014}
and multi-parameter~\cite{Szczykulska2016,Liu2020} estimations. In this theory,
the standard deviation $\delta \delta_0$ is bounded by the inverse of the classical
and quantum Fisher information (CFI and QFI), denoted by $F_{\mathrm{c}}$ and
$F_{\mathrm{q}}$, respectively. In the NPEG-based measurement sheme, the maximized
CFI and the corresponding precision of $\delta_\mathrm{0}$ can be analytically
derived as (see appendix~\ref{CFI} for details)
\begin{equation} \begin{split} \label{C_scaling}
F_{\mathrm{c,opt}} &=\frac{1}{J^2_{\mathrm{eff}}}=\left(\frac{D^{N}(N-1)!}{NU}\right)^2, \\
\delta\delta_\mathrm{0} &= \frac{\delta\Delta_\mathrm{0}}{N}=\frac{U}{D^N (N-1)!},
\end{split}
\end{equation}
indicating the beyond-$1/N$
scaling of the precision on the number of probes.
The maximized $F_{\mathrm{c,opt}}$ is reached at $(\Delta_\mathrm{0},\omega t)=(0,\pi)$,
with the probe state
$\cos\frac{\theta}{2}|\!\!\Uparrow\rangle+\sin\frac{\theta}{2}
|\!\!\Downarrow\rangle$ $(\theta\in[0,\pi])$, which lies on the longitude line connecting
the states $|\pm\rangle$ in the Bloch sphere.
To further check whether the measurement
$\{|\!\!\Downarrow\rangle\langle\Downarrow\!\!|, |\!\!\Uparrow\rangle\langle\Uparrow\!\!|\}$
is optimal, the QFI under the same condition of the maximized CFI can be derived as:
\begin{equation}
F_{\mathrm{q,opt}}=\frac{1}{J^2_{\mathrm{eff}}}=\left(\frac{D^{N}(N-1)!}{NU}\right)^2,
\label{Q_scaling}
\end{equation}
which coincides with $F_{\mathrm{c,opt}}$,
meaning that the MBCIT scheme is an optimal measurement scheme to achieve
the ultimate theoretical precision limit, with a beyond-$1/N$ scaling
behavior~\cite{PhysRevLett.72.3439}.

The scaling in Eqs.~(\ref{C_scaling}) and~(\ref{Q_scaling}) is restricted to
$\Delta_\mathrm{0}=0$, and corresponds to the local precision~\cite{PhysRevX.2.041006},
which at first sight might seem to limit the applicability of the NPEG-based
scheme. The compensate detuning introduced actually circumvents this restriction.
If the compensate detuning can be controlled to high precision, it is always
possible to shift the total detuning to $\Delta_\mathrm{0}=0$ with a fixed
$\Delta_{\mathrm{TBM}}$, and maintain the overall optimal beyond-$1/N$ scaling
across all regions.
 To avoid the high prior information paradox on $\delta_{\mathrm{TBM}}$,
as discussed in Ref.~\cite{Giovannetti2012}, in practice this process needs to be
performed adaptively, such as performing a rough pre-estimation in advance~\cite{Berni2015}.
We will demonstrate with practical examples to show that this prerequisite of
the compensate potential with the high controllability can be fulfilled in
various practical circumstances and the performance of nonzero
$\Delta_0$ will also be thoroughly discussed.

\begin{figure}
\includegraphics[width=8cm]{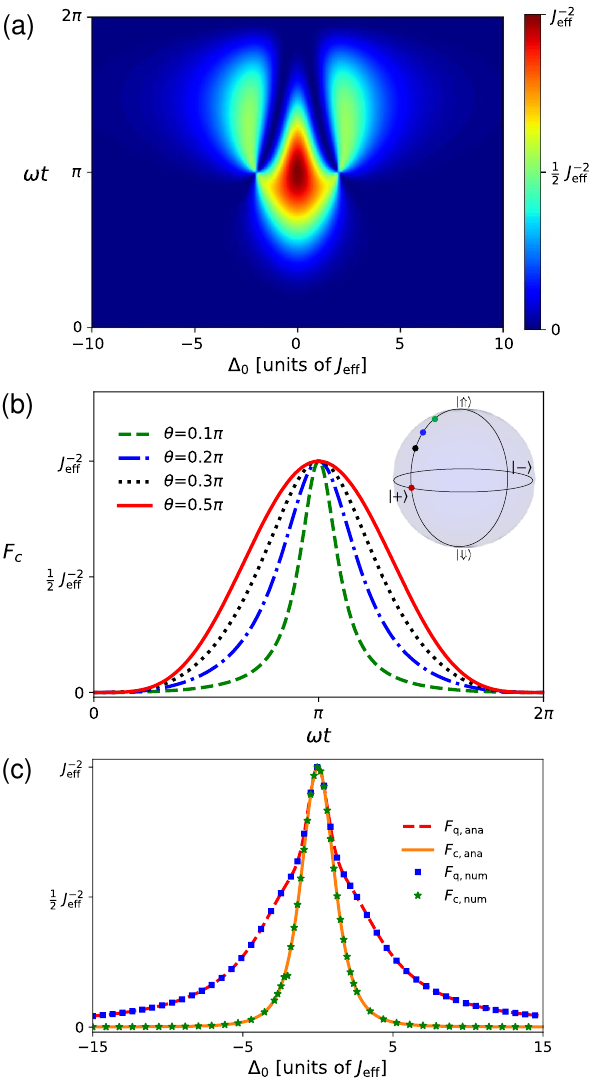}
\caption{(a) The CFI $F_{\mathrm{c}}$ as a function of total detuning $\Delta_{0}$
and $\omega t$ for the initial state $|+\rangle$. (b) Time evolution of $F_{\mathrm{c}}$
for the initial states $\sin(\frac{\theta}{2})|\!\!\Uparrow\rangle+\cos(\frac{\theta}{2})
|\!\!\Downarrow\rangle$ with $\theta=0.1\pi$ (dashed green line), $\theta=0.2\pi$
(dash-dotted blue line), $\theta=0.3\pi$ (dotted black line) and $\theta=0.5\pi$
(solid red line). The insert plot shows the position of these states on the Bloch
sphere. (c) The maximum CFI $F_{\mathrm{c}}$ and QFI $F_{\mathrm{q}}$ as a function
of $\Delta_0$. The dashed red and solid orange lines are the analytical solutions
of $F_{\mathrm{q}}$ and $F_{\mathrm{c}}$ derived via Eq.~(\ref{eq:ham_eff}), and
the blue squares and green stars are the numerical solutions of $F_{\mathrm{q}}$
and $F_{\mathrm{c}}$ calculated with Eq.~(\ref{ham0}), respectively. Here $N$ is
set to be 4.}
\label{fig:fig2}
\end{figure}

Apart from the minimum value of the precision,
the robustness of the achieved precision against imperfect control during measurement
is also an important factor in practical quantum metrology~\cite{Degen2017}.
Figure~\ref{fig:fig2} demonstrates the robustness
against deviations to the optimized rescaled measurement time
$\omega t$ and the total detuning $\Delta_\mathrm{0}$ (Fig.~\ref{fig:fig2}(a)),
as well as against the initial state in terms of the relative phase
$\phi$ (Fig.~\ref{fig:fig2}(b)).
First, the general behavior of the CFI as functions of
$\Delta_\mathrm{0}$ and $\omega t$ is given in Fig.~\ref{fig:fig2}(a)
for the probe state $|+\rangle$.
It is seen that at $(\Delta_\mathrm{0},\omega t)=(0,\pi)$ $F_\mathrm{c}$ reaches
the maximum value, and is most stable with respect to the deviations of the
detuning and the measurement time. Second, Figure.~\ref{fig:fig2}(b) illustrates
that for probe states residing on the longitude line of $\varphi=0$, $F_{\mathrm{c}}$
takes the same maximized value, while the line width of $F_{\mathrm{c}}$ as a
function of $\omega t$ increases as the probe state approaches $|+\rangle$ (also
$|-\rangle$). This indicates that the probe states $|\pm\rangle$ provides the
precision of the best robustness with respect to the measurement time. In
Fig.~\ref{fig:fig2}(c), we further examine $F_{\mathrm{c}}$ and $F_{\mathrm{q}}$
derived from the approximate $H_\mathrm{eff}$, with the results numerically
calculated by $H_\mathrm{org}$. The analytical results based on $H_\mathrm{eff}$
are found to coincide with numerically obtained ones from $H_\mathrm{org}$, which
verifies the validity of the analysis based on $H_\mathrm{eff}$ and confirms the
engineered nonlinear NPEG interaction as a useful quantum resource.
Besides, it also shows that the performance of the scheme is robust
to $\Delta_0$ as in the plot the CFI for $\Delta_0=\pm 0.5$ is around $89.5\%$ of
that for $\Delta_0=0$, indicating that the scheme can be well performed adaptively.

The interrogation time is also an important criterion for the performance of
different measurement schemes~\cite{Giovannetti2001,PhysRevX.8.021022}, thus it
is necessary to evaluate how the NPEG interaction benefits the measurement in
view of time resource. For this purpose, we compare the time cost to reach the
same precision in both linear and nonlinear measurements using the MBCIT process.
The linear MBCIT measurement corresponds to loading only a single probe to the
measurement procedure and eliminating the nonlinear interaction effect,
i.e., taking $N=1$ in the analysis. The ratio of the time cost to reach
the same precision for the linear and the $N$-probe nonlinear MBCIT measurements,
turns out to be  $D^{N-1}(N-1)!$ (see appendix~\ref{Time}),
which indicates that time cost to reach the same precision
also decreases exponentially with $N$.

\section{Application scenarios}
Besides the beyond-$1/N$ scaling, the NPEG-based measurement scheme also enjoys
the advantage of the high flexibility, which allows it to be applied to a
wide range of quantities. In this section, we present several experimental setups
with ultracold atomic ensembles, on which the NPEG-based scheme can be performed
to measure different quantities, such as gravity, magnetic field and its gradients.

\begin{figure}
\includegraphics[width=7.5cm]{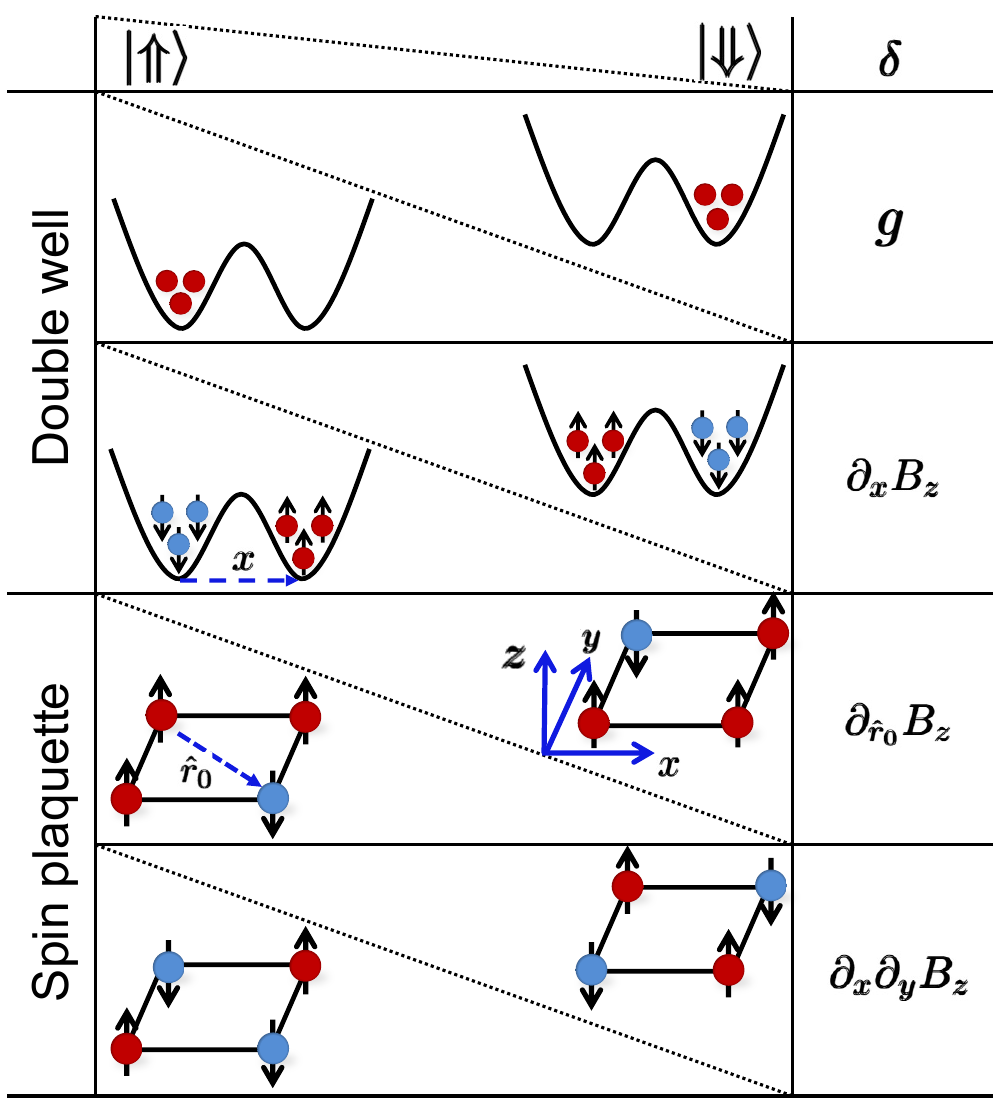}
\caption {Schematics of the practical platforms discussed in the main text
for performing the MBCIT measurements, with the choice of the probe states
and the TBM quantities.}
\label{fig:fig3}
\end{figure}

Our first example explores a system of ultracold atoms confined in a double-well
potential, and various nonlinear interactions have been already discussed and
investigated in this setup~\cite{dudarev2003entanglement,S.F2007,PhysRevLett.100.040401,
PhysRevLett.107.210405,Dai2016}, and the NPEG-based measurement scheme can be
directly applied. As indicated in the first row of Fig.~\ref{fig:fig3}, the NPEG
interaction coupling the two many-body states of atoms confined in the subspace
of all in the left or right wells, can be engineered by the interplay between
the contact interaction and the single-particle hopping. The NPEG-based measurements
of the energy detuning between the two wells can be performed, allowing for
estimations of gravity acceleration along the orientation of the double well.
As illustrated in the second row of Fig.~\ref{fig:fig3}, a second type
of NPEG interaction can be engineered between spinor atoms confined
in the double-well potential~\cite{PhysRevLett.85.3987,PhysRevLett.89.090401},
and couples the states of, for instance,
$|\!\!\Uparrow\rangle=|\!\!\uparrow^{\otimes N},\downarrow^{\otimes N}\rangle$ and
$|\!\!\Downarrow\rangle=|\!\!\downarrow^{\otimes N},\uparrow^{\otimes N}\rangle$,
where each well is occupied by $N$ atoms, and atoms in different wells are
polarized along opposite directions.
Such a probe space is insensitive to the net magnetic field and can be used
for \emph{in-situ} measurement of the linear magnetic gradient
along the orientation of the potential. In both examples, the
highly-controllable lasers forming the double well potential can induce
compensating detuning with required precision.

The NPEG-based measurement can also be applied to spinor atom plaquette,
~\cite{paredes2008minimum,nascimbene2012experimental,dai2017four-body}.
In the plaquette, the nonlinear interactions of the NPEG form
are realized within different energetically resonant subspaces of
the plaquette system, and each such subspace provides a probe space $\mathbb{S}$
for the NPEG-based measurement scheme. The last two rows of Fig.~\ref{fig:fig3}
present two probe spaces capable of measuring the magnetic gradients of different
orders. In the configuration that the spin plaquette lies in the $xy$ plane
and the TBM magnetic field polarized along the $z$-axis, denoted by $B_z(\vec{r})$,
the probe space spanned by
$\{|\!\!\uparrow\downarrow\uparrow\uparrow\rangle,
|\!\!\uparrow\uparrow\uparrow\downarrow\rangle\}$
(indicating the polarization of the four spinors in the plaquette), allows the
\emph{in-situ} measurement of the first-order magnetic gradient $\partial_{\hat{r}_0}B_z$,
with $\hat{r}_0$ along the diagonal line of the plaquette (slashed line in the
third row of Fig.~\ref{fig:fig3}. Turning to the probe space of
$\{|\!\!\uparrow\downarrow\uparrow\downarrow\rangle,
|\!\!\downarrow\uparrow\downarrow\uparrow\rangle\}$, the total spin of the probe
states is zero, and such a probe space provides a good candidate for the \emph{in-situ}
measurement of the second-order magnetic gradient $\partial_{x}\partial_{y} B_z$.
The spin-dependent lattice potential is also readily implemented in the setup,
which can induce the single-particle compensate detunning.

\begin{figure}
\includegraphics[width=8cm]{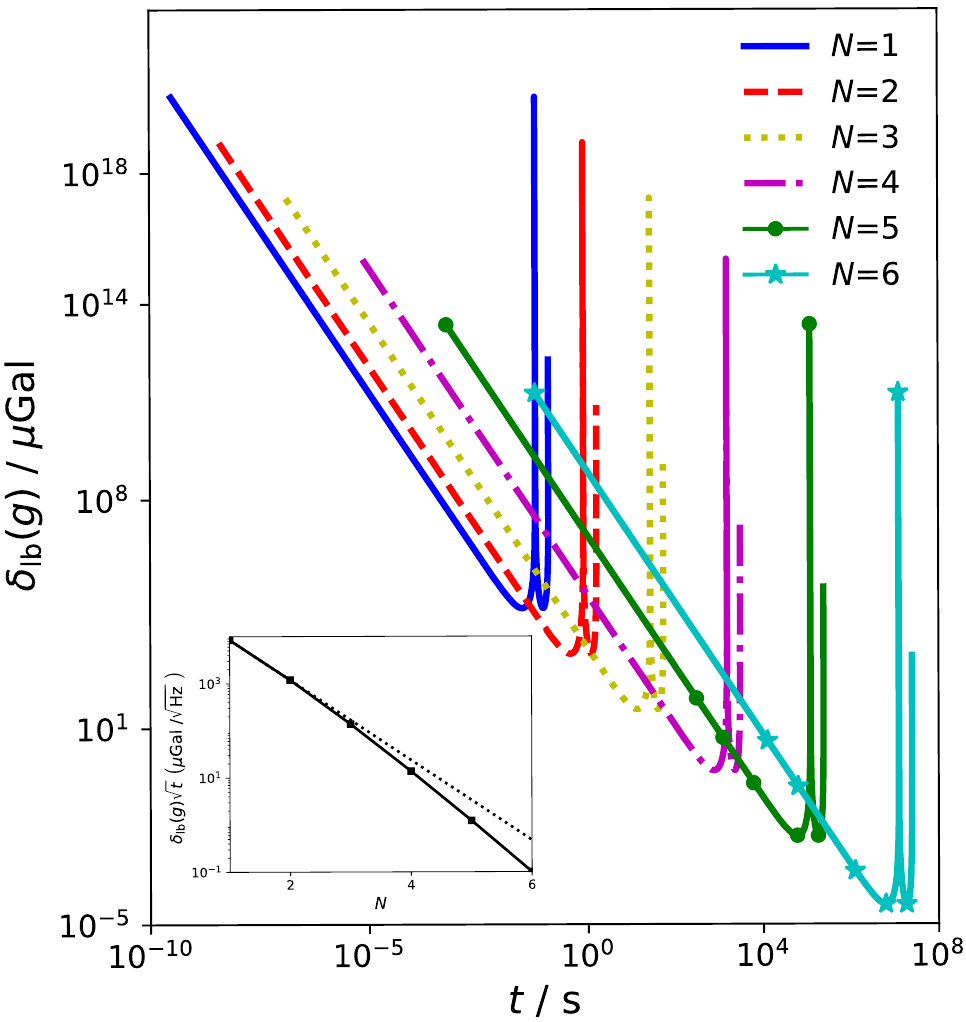}
\caption {The lower-bound uncertainty $\delta_{{\rm lb}}(g)$ as a function of
measurement time $t$ for different particle numbers ($N=1$ to $6$) with $M=10000$
supercells. The parameters are taken from Ref.~\cite{2020Sci_Y}, with the
wavelength of the “short” lattices $\lambda_{s}=767{\rm nm}$, the tunneling
$J=50{\rm Hz}$ and the interaction $U=1251{\rm Hz}$. The insert shows
$\delta_{{\rm lb}}(g)\sqrt{t}$ as a function of $N$, in which the dotted line
presents an exponentially decaying curve.}
\label{fig:fig4}
\end{figure}

We further take the measurement scenario introduced in the first row of
Fig.~\ref{fig:fig3} as an example to quantitatively analyze the performance of
the NPEG-based scheme in the practical measurements. We consider an array of
double wells, each of which is loaded with $N$ ultracold atoms as the probe.
The atoms are subjected to the gravity force and the gravity acceleration $g$
of the Earth is the TBM quantity. In the measurement setup, we consider that
the NPEG-based measurements are performed in parallel for $M=10000$ double-well
supercells. We calculate the lower-bound uncertainty of measured $g$ as a function
of the measurement time $t$ for $N\in [1,6]$, with all relative parameters taken
from the experimental setup in Ref.~\cite{2020Sci_Y}. As shown in the main figure
of Fig.~\ref{fig:fig4}, the lower-bound uncertainty $\delta_{{\rm lb}}(g)$
monotonically decreases with the measurement time, until reaching a minimum.
The minimized $\delta_{{\rm lb}}(g)$ corresponds to the optimized Fisher information
derived in Eqs.~(\ref{C_scaling}) and (\ref{Q_scaling}), and the measurement
time reaching this minimum is just the optimal measurement time. Moreover, as
shown in Fig.~\ref{fig:fig4}, increasing $N$, on the one hand, can exponentially
suppress $\delta_{{\rm lb}}(g)$, and on the other, it increases the measurement
time, which bring in a competition in the practical measurements between minimizing
the uncertainty and shortening the measurement time. Current experiments on the
ultracold atomic dynamics can reach a coherent time over $10$\,s, and setting the
measurement time on this order, one can find that $N=3$ provides the best
lower-bound uncertainty approaching $10\,\mu$Gal, which confirms the NPEG
scheme as a source of high precision measurements. Furthermore, assuming that the
dynamical coherence time of lattice ultracold atomic experiments can approach
the order of $10^3$\,s in the near future, measurements in a double-well array
containing $M=10000$ supercells with each confining $N=4$ bosons provides the
measurement uncertainty below $1\,\mu$Gal. In the insert of Fig.~\ref{fig:fig4},
we further plot $\delta_{{\rm lb}}(g)\sqrt{t}$ as a function of $N$, which is a
joint measure of the performance with respect to the precision and the time
resource. The result illustrates that $\delta_{{\rm lb}}(g)\sqrt{t}$ decreases
(beyond-)exponentially with $N$, which further confirms the beyond-$1/N$
scaling in the joint viewpoint of precision and the time resource.

The highly developed controllability of ultracold atomic ensembles can facilitate the
flexible engineering of the NPEG-type interaction and the compensating detuning with
the required precision. The isolation of atomic ensembles from their environment can
further suppress the noise induced decoherence to a great extent, and guarantees the
long enough coherence time for the proposed measurement~\cite{PhysRevLett.79.3865}.

\section{Summary and discussion}
We have explored the NPEG interaction as a resource for quantum measurement,
and proposed the measurement scheme based on the NPEG-induced MBCIT dynamics.
We have demonstrated that, (i) the NPEG nonlinear interaction manifest itself
as a novel measurement resource, which provides an improved precision limit
with the beyond-$1/N$ scaling of $1/(D^N(N-1)!)$; (ii) the measurement scheme
based on the MBCIT process provides a practical and optimal measurement scheme
to achieve the beyond-$1/N$ scaling.

The NPEG-based measurement scheme not only provides a new mechanism
to invoke nonlinearity and leads to a new type of scaling behaviors, its high
applicability, as illustrated by the ultracold atomic ensemble, magnetic
 molecules~\cite{1980carboxylate,1984Hydrolysis} and nuclear-magnetic-resonance
setups~\cite{PhysRevA.66.044308}, also facilitate the development of next-generation
high-precision apparatuses like gravimeters and magnetometers and benefit both
the fundamental and applied science.

\begin{acknowledgments}
The authors thank L. You, Z.-S. Yuan, Y. Chang, T. Shi and Z. Li for
helpful discussions. The authors particularly acknowledge L. You for carefully
reading through the manuscript and the inspiring suggestions.
This work was supported by the National Natural Science Foundation of China
(Grants Nos. 11625417, 11604107, 11922404, 11727809 and 11805073).
\end{acknowledgments}

\appendix

\section{The effective Hamiltonian}\label{Heff}

\begin{figure}
\includegraphics[width=8.5cm]{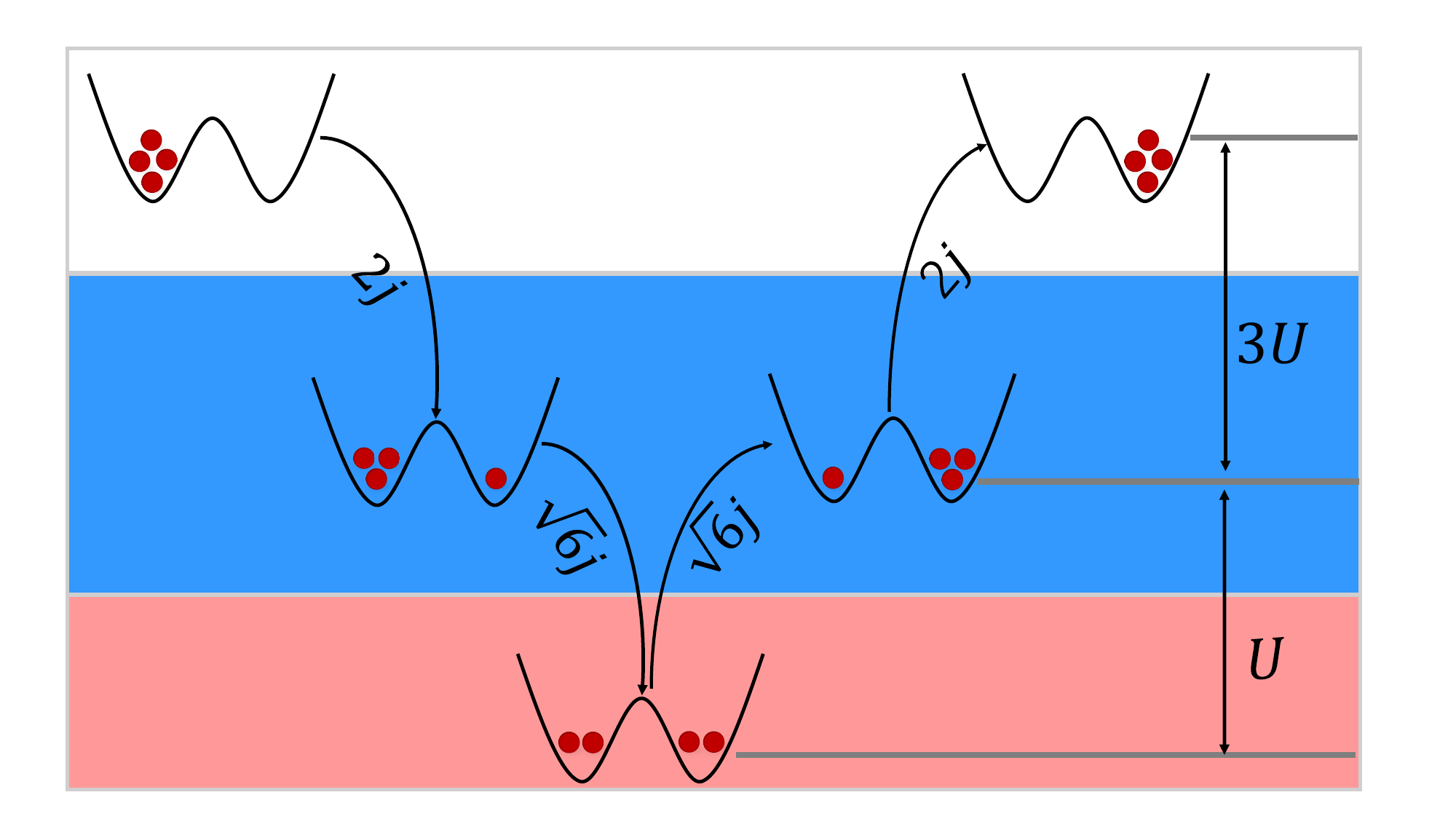}
\caption{ Pathway of the high-order coupling processes in
the high-order perturbation theory for $N=4$.}
\label{fig:apx}
\end{figure}

The system of $N$ Bosons in a double-well trap can be described by a two-mode
single-band Bose-Hubbard Hamiltonian:
\begin{equation} \label{ham01}
H_\mathrm{org}=
\frac{U}{2}(\hat{a}^\dag\hat{a}^\dag\hat{a}\hat{a}+\hat{b}^\dag\hat{b}^\dag
\hat{b}\hat{b})+j(\hat{a}^\dag \hat{b}+\hat{a} \hat{b}^{\dagger})
-\frac{\delta_{\mathrm{0}}}{2}(\hat{n}_a-\hat{n}_b),
\end{equation}
in which $\hat{a}^{(\dag)}/\hat{b}^{(\dag)}$ annihilates (creates) a particle
in the state $|a\rangle/|b\rangle$, and $\hat{n}_{a}=\hat{a}^{\dag}\hat{a}$
($\hat{n}_{b}=\hat{b}^{\dag}\hat{b}$). In the regime of strong interaction
$j\ll U$, the first term of $H_{\mathrm{org}}$ denotes the interaction of
particles, which divides the complete Hilbert space into different energetically
off-resonant subspaces $\{|N-m, m\rangle\!,\,~|m, N-m\rangle\!\}$, while each
subspaces composed of basis states energetically in resonance with each other.
The second term is the one-body hopping term, which builds up higher-order
couplings between basis states in each subspace. The last term denotes the
energy detuning between $|a\rangle$ and $|b\rangle$. Taking the subspace
$\mathbb{S}$ spanned by $\{|N, 0\rangle\!:=|\!\!\Uparrow\rangle,
\,|0, N\rangle\!:=|\!\!\Downarrow\rangle\}$, and the two base vectors can
not be directly coupled with each other, but only by high-order coupling processes,
as demonstrated in Fig.~\ref{fig:apx} for $N=4$. Actually, all vectors
$|N-m, m\rangle$ participate in this high-order coupling processes, and this
effective coupling can be evaluated by high-order perturbation theory
\begin{eqnarray} \label{eq:J_eff1}
J_\mathrm{eff} &=&\frac{\prod_{m=0}^{N-1}\langle m,N-m|\hat{H}_{\mathrm{J}}
|N-m-1,m+1\rangle}{\prod_{m=1}^{N-1}
(\langle 0,N|\hat{H}_\mathrm{int}|N,0\rangle\!-\!\langle m, N\!-\!m|\hat{H}_\mathrm{int}
|N\!-\!m, m\rangle)} \nonumber \\
&=& \frac{NU}{D^{N}(N-1)!},
\end{eqnarray}
where $D=U/j$, $\hat{H}_\mathrm{J}$ and $\hat{H}_\mathrm{int}$ denotes the
hopping and the interaction part of the $H_{\mathrm{org}}$, respectively.
Moreover, the third term of $H_{\mathrm{org}}$ induces the detuning between the
two vectors $|N, 0\rangle$ and $|0, N\rangle$ can be written as
$\Delta_\mathrm{0}= N \delta_{\mathrm{0}}$ and the engineered Hamiltonian becomes
\begin{equation}
\label{eq:heff}
H_\mathrm{eff}\!=\!J_\mathrm{eff}\left(|\!\!\Uparrow\rangle\langle\Downarrow\!\!|
+|\!\!\Downarrow\rangle\langle\Uparrow\!\!|\right)
-\frac{1}{2}\Delta_\mathrm{0}\left(|\!\!\Uparrow\rangle\langle\Uparrow\!\!|
-|\!\!\Downarrow\rangle\langle\Downarrow\!\!|\right).
\end{equation}

\section{Calculations of the CFI}\label{CFI}

The temporal evolution of the system during the effective Hamiltonian in
Eq.~(\ref{eq:heff}) for a general probe state of
$\cos\frac{\theta}{2}|\!\!\Uparrow\rangle+e^{i\varphi}\sin\frac{\theta}{2}
|\!\!\Downarrow\rangle$ ($\theta\in[0,\pi]$, $\varphi\in [0,2\pi]$) is given by
$|\psi(t)\rangle=C_{\Uparrow}(t)|\Uparrow\rangle+C_{\Downarrow}(t)|\Downarrow\rangle$,
where
\begin{eqnarray*}
& & C_{\Uparrow}(t) \\
&=& \cos\!\left(\frac{\omega t}{2}\right)
\!\cos\!\left(\frac{\theta}{2}\right)\!+\! 2\mathrm{sgn}(J_{\mathrm{eff}})
\frac{\sin(\frac{\theta}{2})\sin(\frac{\omega t}{2})\sin\varphi}{\sqrt{4+\Gamma^2}} \\
& & +i\!\left\{\frac{\mathrm{sgn}(J_{\mathrm{eff}})\sin^2(\frac{\omega
t}{2})}{4+\Gamma^2}\!\left[\Gamma\!\cos\!\left(\frac{\theta}{2}\right)\!-\!
2\sin\!\left(\frac{\theta}{2}\right)\cos\varphi\!\right]\!\right\},
\end{eqnarray*}
that is to say, the probability of the measurement $\{|\!\!\Downarrow\rangle
\langle\Downarrow\!\!|,|\!\!\Uparrow\rangle\langle\Uparrow\!\!|\}$ reads
\begin{eqnarray}
\label{proba}
P_{\Uparrow}(t) &\!=\!&\left[\!\cos\!\left(\!\frac{\omega t}{2}\!\right)
\!\cos\!\left(\!\frac{\theta}{2}\!\right)\!\!+\!\!2\,\mathrm{sgn}(J_{\mathrm{eff}})
\frac{\sin(\frac{\theta}{2})\sin(\frac{\omega t}{2})\sin\varphi}{\sqrt{4+\Gamma^2}}
\right]^2\!\! \nonumber \\
& & +\frac{\sin^2(\frac{\omega t}{2})}{4+\Gamma^2}
\left[\Gamma\cos\left(\frac{\theta}{2}\right)-2\sin\left(\frac{\theta}{2}\right)\cos\varphi\right]^2,
\end{eqnarray}
and
\begin{equation}
P_{\Downarrow}(t)=1-P_{\Uparrow}(t),
\end{equation}
with $\omega=\sqrt{4J_\mathrm{eff}^2+\Delta_\mathrm{0}^2}$ and
$\Gamma=\Delta_\mathrm{0}/J_\mathrm{eff}$.

It is generally known that the classical Fisher information $F_{\mathrm{c}}$ can
be obtained via $F_{\mathrm{c}}=\sum_i(\partial_x p_i)^2/p_i$ with $\{p_i\}$ a
set of probability distribution. For our system,
\begin{equation}\label{F_c_form}
F_{\mathrm{c}}=\frac{(\partial_{\Delta_{\mathrm{0}}} P_{\mathrm{\Uparrow}})^2}{P_{\mathrm{\Uparrow}}}+\frac{(\partial_{\Delta_{\mathrm{0}}} P_{\mathrm{\Downarrow}})^2}{P_{\mathrm{\Downarrow}}}.
\end{equation}

Substitute formula Eq.(\ref{proba}) into formula Eq.(\ref{F_c_form}), we can
find that $F_\mathrm{c}$ is a function consisting of four parameters
$\theta$, $\varphi$, $\omega t$ and $\Delta_\mathrm{0}$ and could be written as
\begin{equation}
\label{F_C}
F_\mathrm{c}(\theta,\varphi,\omega t,\Delta_\mathrm{0})
= \frac{F_\mathrm{C1}(\theta,\varphi,\omega t,\Delta_\mathrm{0})}
{F_\mathrm{C2}(\theta,\varphi,\omega t,\Delta_\mathrm{0})}.
\end{equation}
Here  $F_\mathrm{C1}(\theta,\varphi,\omega t,\Delta_\mathrm{0})$ and
$F_\mathrm{C2}(\theta,\varphi,\omega t,\Delta_\mathrm{0})$ are the numerator
and denominator of $F_\mathrm{c}$, respectively, and could be expressed as
\begin{eqnarray}
& & F_\mathrm{C1}(\theta,\varphi,\omega t,\Delta_\mathrm{0}) \nonumber \\
&=& 4\Big\{2J_\mathrm{eff}\Gamma\cos\theta[2\cos(\omega t)-2+(\omega t)\sin(\omega t)]
\nonumber \\
& & +\sin\theta\left[J_\mathrm{eff}\cos\varphi(\Gamma^2-4)(\cos(\omega t)\!-\!1)
\!+\!\Gamma^2(\omega t)\sin(\omega t)\right] \nonumber \\
& & +\Gamma\sqrt{J_\mathrm{eff}^2(4+\Gamma^2)}[\sin(\omega t)-\omega t
\cos(\omega t)]\sin\varphi)\Big\}^2\!\!,
\end{eqnarray}
and
\begin{eqnarray}
& & F_\mathrm{C2}(\theta,\varphi,\omega t,\Delta_\mathrm{0}) \nonumber \\
&=& J_\mathrm{eff}^2(4+\Gamma^2)^2\Bigg\{J^2_\mathrm{eff}(4+\Gamma^2)^2 \nonumber \\
& & \!-\!\Big[J_\mathrm{eff}(\Gamma^2+4\cos(\omega t))\!\cos\theta
\!-\!4J_\mathrm{eff}\Gamma\cos\varphi\sin\theta\sin^2\!\left(\frac{\omega t}{2}\right) \nonumber \\
& & +2\sqrt{J_\mathrm{eff}^2(4+\Gamma^2)}\sin\theta\sin(\omega t)\sin\varphi\Big]^2\Bigg\}.
\end{eqnarray}
For the initial state $|+\rangle$, which is required for MBCIT, $F_\mathrm{c}$
evolves over time as follows
\begin{eqnarray*}\label{F_C1}
& & F_\mathrm{c}(\theta=\frac{\pi}{2},\varphi=0,\omega t,\Delta_\mathrm{0})  \\
&=& \frac{16\sin^2(\frac{\omega t}{2})(-\Gamma^2\omega t\cos(\frac{\omega t}{2})
+(\Gamma^2-4)\sin(\frac{\omega t}{2}))^2}{J_\mathrm{eff}^2(4+\Gamma^2)^2
(16 + \Gamma^4 - 2\Gamma^2(\cos(2\omega t)-4\cos(\omega t)-1))}.
\end{eqnarray*}
Under this initial state constraint, if and only if $\omega t=\pi$ and
$\Delta_\mathrm{0}=0$, the value of $F_\mathrm{c}$ could attain the maximum.
Remove this initial state $|+\rangle$ constraint for the optimal condition
$\omega t=\pi$ and $\Delta_\mathrm{0}=0$, i.e. for a general probe state $\cos\frac{\theta}{2}|\!\!\Uparrow\rangle+e^{i\varphi}\sin\frac{\theta}{2}
|\!\!\Downarrow\rangle$ ($\theta\in[0,\pi]$, $\varphi\in [0,2\pi]$), $F_\mathrm{c}$
can be written as follows
\begin{equation}
F_\mathrm{c}(\theta,\varphi,\omega t=\pi,\Delta_\mathrm{0}=0)
=\frac{\cos^2\varphi}{J_\mathrm{eff}^2},
\end{equation}
which means that the value of $F_\mathrm{c}$ can reach the maximum under
$\varphi=0~ \& ~\varphi=\pi$ and independent on $\theta$. Then we choose
$\varphi=0$ and $\Delta_\mathrm{0}=0$ and can obtain the evolution of
$F_\mathrm{c}$ under different $\theta$
\begin{equation}
F_\mathrm{c}(\theta,\varphi=0,\omega t,\Delta_\mathrm{0}=0)
=\frac{[\cos(\omega t)-1]^2\sin^2\theta}{4J_\mathrm{eff}^2
[1-\cos^2\theta\cos^2(\omega t)]},
\end{equation}
the robustness of $F_\mathrm{c}$ to the measurement time depends on $\theta$
and could attain the maximum under $\theta=\pi/2$.

For the above three optimal parameters, i.e. $\theta=\pi/2, \varphi=0,\omega t=\pi$,
$F_\mathrm{c}$ can be written as follows
\begin{equation}
F_\mathrm{c}=\frac{16}{J_\mathrm{eff}^2(4+\Gamma^2)^2},
\end{equation}
and in this optimal parameters, the QFI is calculated as
\begin{equation}
F_{\mathrm{q}}=\frac{16}{J_\mathrm{eff}^2(4+\Gamma^2)^2}+\frac{\Gamma^4\pi^2}
{J_\mathrm{eff}^2(4+\Gamma^2)^3}.
\end{equation}
Here $F_{\mathrm{q}}\geq F_{\mathrm{c}}$, if and only if $\Delta_\mathrm{0}=0$,
\begin{equation}
F_{\mathrm{c,opt}}=F_{\mathrm{q,opt}}=\frac{1}{J_{\mathrm{eff}^2}}
=\left(\frac{D^{N}(N-1)!}{NU}\right)^2.
\end{equation}
The establishment of the equal sign reflects that the MBCIT measurement scheme
is optimal.

According to the Quantum Cram\'{e}r-Rao bound~\cite{Helstrom1976,Holevo1982},
\begin{equation}
\delta \Delta_{0}\geq \frac{1}{\sqrt{\nu F_{\mathrm{c}}}}
\geq \frac{1}{\sqrt{\nu F_{\mathrm{q}}}},
\end{equation}
where $\nu$ is the repetition of experiments. Under the optimized condition,
\begin{equation}
\delta \Delta_{0}=|J_{\mathrm{eff}}|=\frac{NU}{D^{N}(N-1)!}.
\end{equation}
Since $\Delta_{\mathrm{0}}=N\delta_{\mathrm{0}}$,
\begin{equation}
\begin{split}
& \delta\delta_\mathrm{0}=\frac{\delta\Delta_\mathrm{0}}{N}=\frac{U}{D^N(N-1)!}.
\end{split}
\end{equation}

\section{Time consumption}\label{Time}

In order to show the advantages of this nonlinear MBCIT scheme in the time
resource consumption, we compare the time cost to reach the same precision
in the nonlinear and linear MBCIT measurements. For the linear MBCIT measurement,
which corresponds to the case loading $N$ sing-body probes to the measurement
procedure and repeat $m_\mathrm{1}$ measurements, the precision and the time
resource consumption could be written as
\begin{eqnarray}\label{uncer_1}
\delta_\mathrm{1}&=&\frac{|j|}{\sqrt{Nm_\mathrm{1}}}, \\
\tau_\mathrm{1}&=&\frac{m_\mathrm{1}\pi}{|2j|}.
\end{eqnarray}
As for the nonlinear MBCIT measurement, which corresponds to the case loading
a $N$-body probe to the measurement procedure and repeat $m_\mathrm{2}$
measurements, the precision and the time resource consumption could be written as
\begin{eqnarray}\label{uncer_2}
\delta_\mathrm{2}&=&\frac{|J_\mathrm{eff}|}{N \sqrt{m_\mathrm{2}}}, \\
\tau_\mathrm{2}&=&\frac{m_\mathrm{2}\pi}{|2J_\mathrm{eff}|}.
\end{eqnarray}
For the same precision, $i.e.$ $\delta_\mathrm{1}=\delta_\mathrm{2}$, the ratio
of the time cost turns out to be:
\begin{equation}
\label{time com}
\frac{\tau_\mathrm{2}}{\tau_\mathrm{1}}=\frac{1}{D^{N-1}(N-1)!}.
\end{equation}

\bibliographystyle{apsrev4-1}

\end{document}